\documentstyle[12pt]{article}

\newcommand\ba{\begin{array}}
\newcommand\ea{\end{array}}
\newcommand\ben{\begin{equation}}
\newcommand\een{\end{equation}}
\newcommand\bea{\begin{eqnarray}}
\newcommand\eea{\end{eqnarray}}

\def\be{\beta}
\def\ga{\gamma}

\def\si{\sigma}

\def\vev#1{\langle#1\rangle}

\title{
{\normalsize
\begin{flushright}
DAMTP--94--56\\
SUSX--TP--94--72\\
hep-th/9410094\\
\end{flushright}}
\vspace{1 cm}
Statistical Properties of Strings}
\author{M. Hindmarsh$^{(a)}$ and K. Strobl$^{(b)}$}

\begin{document}
\maketitle

\vskip 12pt
\begin{center}
{\it (a) School of Mathematical and Physical Sciences\\
  University of Sussex\\
  Brighton BN1 9QH\\
  UK}\\
\vskip 12pt
 {\it  (b) Dept.~of Applied Mathematics and Theoretical Physics\\
          Silver Street\\
          Cambridge CB3 9EW\\
          UK }
\end{center}
\vskip 12pt
\begin{abstract}
We investigate numerically the configurational statistics of strings.
The algorithm models an ensemble of
global $U(1)$ cosmic strings, or equivalently vortices in superfluid
$^4$He.  We
use a new method which avoids the specification of boundary conditions
on the lattice. We therefore do not have
the artificial distinction between short and long string loops or
a `second phase' in the string network statistics
associated with strings winding around a toroidal lattice. Our lattice is
also tetrahedral, which avoids ambiguities associated with the cubic
lattices of previous work.  We find that the percentage of infinite
string is somewhat lower than on cubic lattices, 63\% instead of 80\%.
We also investigate the Hagedorn transition, at which infinite strings
percolate, controlling the string
density by rendering one of the equilibrium states more probable.
We measure the percolation threshold, the critical exponent associated
with the divergence of a suitably defined susceptibility of the string
loops, and that associated with the divergence of the correlation
length.
\end{abstract}
\newpage

Line defects are formed after a phase transition if the manifold
of equilibrium states $M$ (the vacuum manifold) is not simply connected
\cite{Kibble}. They are studied theoretically in the context of field
theories in the early Universe under the name of cosmic strings,
but also exist in the laboratory in the form of superfluid
vortices, superconductor flux tubes, dislocations, and
line disclinations in nematic liquid crystals.  The
interest in the formation of defects in cosmological phase transitions
has been reflected in laboratory experiments studying the formation
and evolution of defects in nematic liquid crystals \cite{Che+90,Bowick}
and He-II \cite{McLin}.

In this paper we look at the configurational statistical of $U(1)$
strings, where $M$ is a circle.  The properties of ensembles of
1-dimensional objects have importance in many problems in physics:
polymer science \cite{deG}; dislocation melting \cite{Disloc}; the
liquid-gas transition \cite{Riv}; and the Hagedorn transitions in
effective \cite{Hag} and fundamental \cite{BowCha,MitTur} theories of
strings at high temperature.  The notion that the entropy
associated with the strings plays a key role in the superfluid
phase transition goes back to Feynman and Cohen \cite{FeyCoh}, and
has been expressed in a field theory context by Copeland et
al.~\cite{Cop+}.
Strings are usually modelled by random
walks, either Brownian or self-avoiding.  A self-avoiding walk (SAW)
models an excluded volume effect, and is known to apply well to
polymers \cite{deG}.  However, it is not clear that either kind of
walk represents the configurational statistics of cosmic strings or
superfluid vortices, for there are long-range
interactions which could change the Hausdorff dimension. In fact,
we find that at low density strings are self-{\it seeking}:
that is, their fractal dimension is higher than 2.

The formation of $U(1)$ strings has been studied by several workers in
simulations on cubic lattices \cite{VV,V91,Allega}. Kibble \cite{KibZ2}
has studied $SU(2)$ strings, and others have studied composite defects
such as monopoles joined by strings \cite{Compo}.
Much of this
work has been done on finite size lattices with periodic boundary
conditions, which suffers from boundary effects that have nothing to do
with the true ensemble \cite{Allega,Aus+}.
We propose a formalism which can simulate
an arbitrarily large lattice, without considering more points than necessary
to follow a single string. This allows us to follow string statistics to
string lengths of $10^5$ lattice units, without prohibitive memory
requirements. This is a necessity for the accurate determination of the
statistical measures of the model: the fraction of the total string
mass contained in infinite strings; the percolation threshold at which
this fraction vanishes; the exponent for divergence of the
susceptibility at this threshold; and the Hausdorff dimension of
the strings.

Before we describe the numeric procedure in detail,
we will motivate our measurements in the following section and explain
what results one expects intuitively.

\section{String Statistics and the Scaling Hypothesis}

One might expect the network of cosmic strings to have the
statistical properties of a self-avoiding random walk.
A SAW builds up an
excluded volume as it follows its path, which is, in a statistical
sense, spherically symmetric and clustered around the origin. The SAW
therefore has a stronger tendency to move away from the
origin than the Brownian walk, which is allowed to intersect itself
arbitrarily often. This property is expressed in the fractal dimension $D$
of the walk, which is the exponent relating the average string length $l$
between two points on the same string to their relative distance $R$ by
\begin{equation}
\l\propto R^{D}\mbox\ .
\label{eq:dimension}
\end{equation}
It is well known that the dimension for a Brownian walk is $D=2$ and for a
self-avoiding random walk $D=1/\nu=1/(0.5879\pm0.0005)$
(see ref.~\cite{Sokal} and
references therein for a summary of different methods used to obtain that
result).
However, the original string formation simulations \cite{VV,V91} are
consistent with $D=2$: this is because they simulated a dense
string network.
A single string, as we trace out its path,
experiences a repulsion from all of the segments of other strings, which
do not have any statistical bias towards the origin.
Therefore the repulsion
from the forbidden volume will also have no directional
bias. Thus
the fractal dimension of the string is expected to be two. In polymer
physics, this effect has been known for some time to occur in dense
solution of polymers, while a polymer in a dilute solution exhibits the
statistics of a self-avoiding random walk \cite{deG}.
In a statistical sense, the network
of cosmic strings was argued to be equivalent to a dense network of
polymers \cite{Scherrer}.

We can introduce the scaling hypothesis in order to estimate a few other
properties of the string network. Scale invariance means that the string
network looks the same on all scales in terms of statistical properties.
In fact Brownian random walks are scale invariant. With
this hypothesis, the expected
distribution of closed loops can be derived \cite{Vil}.
 From dimensional arguments, the number of closed loops with size from $R$ to
$R+dR$ per unit volume can be written as
\[
dn=\frac{dR}{R^4}f\left(\frac{R}{\xi}\right)\mbox\ .
\]
If the system is scale invariant, the distribution should be independent of the
correlation length $\xi$, and one expects
\begin{equation}
dn\propto R^{-4}dR\mbox\ .
\label{eq:dn}
\end{equation}
The length distribution of loops for strings with a fractal dimension of
$D$ is therefore
\begin{equation}
dn\propto l^{-b}dl,
\label{eq:lengthdistr}
\end{equation}
with $b=1+3/D$.
It was originally expected \cite{VV} that from scale invariance there
should be no infinite strings. This turned out not to be the case,
since infinite strings
can still look statistically the same on all scales: a Brownian walk
is scale invariant and has a non-zero probability not to return to the origin
in $d>2$ dimensions.
The origin of the scale invariance of the string network seems
to be connected with the
absence of long-range correlations in the order parameter \cite{VV}.
However, scale invariance does not necessarily imply that the network
is Brownian as originally stated: it can be seen above that one
does not need $D=2$ in order to have a scale-free distribution of
loop sizes $R$.

Vachaspati \cite{V91} devised an algorithm that induced correlations
in the order parameter by lifting the degeneracy of the manifold of
equilibrium states.  This had the effect of reducing the
density of string, but {\it increasing\ }the dimension $D$, which
argues against the identification of strings with SAWs.  He found that
there was a critical density below which there were no ``infinite''
strings, which for a finite size lattice of dimensions $N^3$ are all
strings longer than $kN^2$, with $k=O(1)$ \footnote{The measured
percentage of mass in infinite strings is rather insensitive to the
particular choice of $k$ \cite{V91}. However, string loops
which wind around the
toroidal lattice create difficulties in the interpretation of the variables
in Eq.\ (\ref{eq:dimension}). This explains why our measurements disagree
with \cite{Allega}, who considered winding loops to be a
new long-string phase with different scaling exponents.}.
In the low density phase there is a scale $c$
which appears in the loop length distribution,
\ben
dn = a l^{-b} e^{-cl} dl,
\label{eq:lengthdistr2}
\een
as a cut-off. As the critical density is approached from below, $c\to 0$,
and the mean square fluctuation in the loop length
\ben
S = \vev{l^2} - \vev{l}^2,
\label{eq:S}
\een
diverges (see Figure 7).
This divergence signals a phase transition, called the Hagedorn
transition, which arises from a density of states which is exponentially
increasing with energy.
As mentioned in the introduction, it has been implicated in
many branches of physics, although Vachaspati's algorithm most directly
models the symmetry-breaking phase transition in $U(1)$ scalar field
theory or the normal-superfluid transition in liquid helium.  Previous
studies \cite{Cop+}  assumed from the
outset that the strings are Brownian, and tried to model the true string
ensemble by adding in interactions.  Vachaspati's algorithm enables
us to measure directly the string statistics such as the critical
density, the dimension, and the critical exponents,
and to test the validity of the hypothesis of scale invariance.

\section{Choosing a lattice}

The general idea of such simulations is to put a grid onto space, with the
average nearest neighbour distance representing the correlation length
of the order parameter.  Now, to every vertex of the lattice,
we can assign a random value to the
order parameter  (here a complex scalar field), subject to the constraint
that it lie in the manifold of minima of the free energy $M$.
The correlation length $\xi$ is certainly less than or
equal to the cosmological horizon distance $ct$ at the time of the phase
transition.  In fact, in a second order transition, it is just the
Compton wavelength of the scalar particle, while in a first order
transition, such as occurs in a liquid crystal, it is interpreted
as the average bubble separation.  Thus in the second order case we are
simulating the thermal ensemble, and in the first order case we are
simulating the nucleation of bubbles with random phases.
Furthermore, following Vachaspati and Vilenkin \cite{VV}, we
discretize $M$ by allowing
the phases of the scalar field at the
vertices to take only one of the three values:
$\theta=0,\ 2\pi/3,\ 4\pi/3$.
A string passes through the face of the space-grid, if the scalar field rotates
through $2\pi$ when traced around the edges of that face. In the process of
tracing the field between two vertices of the lattice, we assume that the
interpolation between the two values assigned to the vertices is such that the
change of $\theta$ is minimal -- the so-called geodesic rule
\cite{Kibble,RudSri,Hin+}. If we apply this rule, a face is penetrated exactly
if all the three possible values of $\theta$ occur on the face in question. It
is easy to convince oneself that for a tetrahedron this has to be the case
for either two or none of the four faces, thus the geodesic rule ensures
conservation of the `string flux' and no string can be terminated except by
building a closed loop. Discretizing $M$ is numerically sensible,
for it allows the use of integer arithmetic, but it does change the
probability that a string passes through a face \cite{LeePro}.  It
may also change the percolation probability, and possibly even
the critical exponents. This is something we plan to investigate in
future.

Using this construction to simulate string formation
on a cubic lattice leaves ambiguities in the definition
of the network, since some unit cells will be penetrated by two strings.
In this case we would have to assign
connections of the two incoming and two outgoing segments randomly.
It is not obvious {\it a priori} with what probabilities to make
these assignments.  Vachaspati and Vilenkin chose equal probabilities,
and found a long string density significantly higher than ours
(see below).  It seems likely that connecting strings in adjacent faces
more often will make loops more common.

A tetrahedral lattice does not suffer from this problem.
One has to be quite careful to choose a tetrahedal
lattice which
produces rotationally invariant string statistics.
No tetrahedral lattice which is only a subdivision of a cubic lattice can
achieve that.
We made measurements with a cubic lattice split up into the same set of six
tetrahedra, as well as with a lattice on which the subdivision of the cubic
cell into tetrahedra
was programmed to happen randomly. In either case  one body
diagonal is always
preferred, for it occurs either in all of the cubes or in none of them.
Isotropy of the string distribution was measured by an inertia
tensor. Instead of measuring only $R^2$, as is necessary
to measure the fractal dimension and test the validity of
Eq.\ (\ref{eq:dimension}), we measured
$I_{ij}(l)=\vev{R_i(l)R_j(l)}$.
In the case of any subdivided cubic lattice this tensor turned out
not to be proportional to the unit matrix $\delta_{ij}$, as one
must have for a statistically isotropic string distribution.

Among the tetrahedral lattices we tested, only the one which is the dual
lattice to a tetrakaidekahedral lattice proved to have this
property.\footnote{A random
lattice with a proper triangulation would also satisfy
the requirement of statistical isotropy, but such a simulation would
be much more computationally intensive. Also, the lattice spacings would
differ to a larger extent, which would represent a first order phase transition
to produce the strings after (randomly distributed) bubble collisions.
We don't expect any different results on a random lattice.}
This is in effect a tetrahedral lattice
with a set of vertices building up a body-centered cubic lattice. The
unit cell of the tetrahedral lattice is shown in Figure 1 with the twelve
equivalent tetrahedra it consists of. Another useful feature of this lattice
is \cite{Scherrer} that the edges of all the tetrahedra are nearly equally
long and therefore can be said to represent a given correlation length $\xi$
rather well. There are only two different edge lengths with a length
ratio $2:\sqrt{3}$. This is also reflected in
the fact that the first Brillouin zone of the body-centered cubic lattice,
which
builds up the tetrakaidekahedral lattice, is nearly spherical. The dual
lattice to the tetrahedral lattice is shown in Figure 2. The possible string
paths lie on the edges of the dual lattice, while the field values are
assigned to the center of each tetrakaidekahedron.

In our measurements, we started out by considering a single tetrahedron being
penetrated by a string. We then followed the string along whichever
path it was forced to take by the random field value assigned to the new vertex
of the next tetrahedron penetrated.
At every step we have to check whether we have
been at the presently relevant space-point already.
Since 24 tetrahedra are adjacent to
each lattice point on
the body-centered cubic lattice, the field value at a single lattice
point can be required several times while tracing a string.
To be able to recall these values without wasting memory on lattice points
which
are never needed, we
created a linked list of structures, whose elements carried the essential
information of the already used lattice points.
The advantage lies in the realization that lattice points which are not
adjacent to the string under consideration are irrelevant, and that the way the
string is constructed itself ensures that the string ``knows" about the string
density around it. This is the crucial improvement in our treatment.
The computational effort to check the whole list of
structures at every step increases
proportionally to the square of the maximum length of the strings.
We created a linked list of such linked lists, and searched in the list
containing only certain elements, e.g.\ only elements within a certain
distance interval from the origin. This is effectively the same as using
a very undersized hash table with collision resolution by chaining
\cite{hashtable}. While this work is being completed, the authors plan
for future work to
use an oversized array as a hash table, allowing for a ``nearly unique"
hash function which, compared to the simple distance-mapping we used here,
produces many fewer data collisions and will reduce the search time to
the order of the time it takes to {write down the new step}, so that
the computational effort of tracing a string will be directly proportional
to the string length.
However, even our undersized hash table speeded up
the computation by a factor of about 700 compared to simple chaining
and were able to trace strings up
to lengths of 100,000 correlation lengths, and to do statistics over 10,000
of these strings, in about 24 hours on a HP 715/50. In fact, with our
algorithm the length of traceable strings is limited only by the
memory required to store {\it one} string.
In comparison with previous calculations this allows very
good statistics on long strings without running out of memory.
If one creates a finite size lattice and assigns
field values to all the points on the lattice before tracing out the strings
(as it was done in all the previous work on this subject), the longest strings
occuring (regardless of the boundary conditions) are expected to be
about $N^2$ correlation lengths long, if $N^3$ is the number of lattice points.
In order to
achieve the same maximum length, one would therefore have to use a
$300^3$ lattice.
If one counts
strings near $N^2$ correlation lengths on a $N^3$ lattice, one
either gets spurious effects from the boundary conditions \cite{Allega},
or one faces problems of biased statistics.\footnote{
For example, with absorbing boundaries we count only the very crumpled strings
among
the very long ones, and the measured Hausdorff dimension of the long strings
will therefore be larger than the actual one}

\section{Results for Equiprobable Field Values}

We first set about measuring the statistics for equiprobable field
values,
that is, that all three values of the field had the same probability
at each vertex.   This models the statistics of the string network
at formation.
In Figure 3 we plotted  $R^2$ against
$l$, which should be exactly linear in the Brownian case (and therefore it
should have slope one on the logarithmic plot). In fact we measure for the
fractal dimension of the $U(1)$ strings
\begin{equation}
D=2.01\pm 0.01\mbox\ ,
\label{eq:nu}
\end{equation}
which indicates no repulsion from the origin, but perfect Brownian statistics
instead. We detect no deviations from a perfect scaling behaviour, and as soon
as the length of string under consideration is an order of magnitude longer
than the lattice spacing, this coefficient is completely independent of the
length of the strings considered.

There is a subtlety to discuss: even though the scaling
might be Brownian, this equation
cannot be understood as ``after $l$ steps one is likely to be at the distance
$R$ from the origin", since there is a certain probability that the walk has
terminated by then, which does not happen to a Brownian walk. We obviously
only average over the strings that have survived up to the distance $l$.
Of course, for Brownian walks, if we average them only over the ensemble
that has not revisited the origin until it got to a length $l$ we still
get a fractal dimension $D=2$, because $D$ is measured in the asymptotically
large $l$-limit, where the probability of return to the origin tends to
zero {\em for all} Brownian walks, whether they have revisited the origin
in the past or not. A Brownian walk is Markovian and does not remember its
past, so every walk will revisit the origin only a finite number of times,
if the fraction of walks that never come back is nonzero, which
is trivially true. The probability of return to the
origin, however, depends very much on the microscopic details of the lattice,
for instance on whether we allow ``full turns" or not. For example, as
a test we
produced our random walks the same way as we produced the cosmic strings,
but did not store the phases if we revisited a given lattice site. This
disallows full turns, but does not force the walk to avoid itself.
The total walk mass in loops was then only
$\approx 9\%$. Conversely, since the coordination number of the
lattice on which our
strings live is four, a quarter of all the walk
mass would be in loops of only the smallest possible size (length 2), if
we allow immediate returns on the random walk.

We measure that $(63.3 \pm 1.0)$\% of the total string mass is in infinite
strings, compared to the Vachaspati and Vilenkin result of about 80\%
\cite{VV}. This is an indication that the random assignment of string
elements in cubes that are crossed by more than one string is statistically
misleading, and that strings prefer to crumple up to a larger extent than
is assumed by such a random assignment. Unfortunately this is not a
quantitatively hard result (and neither was the one of \cite{VV}), for
reasons that are mentioned above. Some of the
difference in our results may come therefore from the lower coordination
number of our dual lattice, so that short loops are
more probable than on a cubic lattice (4/125 of all the string mass is
in four-step loops on a cubic lattice, whereas on our lattice this ratio
is 1/27).
Another reason why the mass in small loops affects our results is that
most of the
string mass that is in loops, is in small loops, and our lattice allows
loop diameters that are $1/\sqrt{3}$ times smaller (in units of
correlation lengths) than the smallest loops allowed in ref.~\cite{VV}.
In our formalism it is easy to convince oneself that these 63.3\% are simply
equal to the ratio of strings which have not terminated and run into themselves
after the maximum number of steps. It should be mentioned that at string
lengths of 100,000 $\xi$ this is very
insensitive to the cutoff-length of our calculation.
The discretization of
the symmetry group and of 3-space is certainly introducing bigger inaccuracies
than the cutoff at that length.

To measure the fraction of loops within a certain length-interval, i.e.\ the
exponent in Eq. (\ref{eq:lengthdistr}), we use the linear fit in Figure 4,
which
is a logarithmic plot of the density of string elements in loops of a certain
length interval. From (\ref{eq:lengthdistr}) this relation should be
\begin{equation}
dN\propto l^{1-b}dl\mbox\ ,
\label{eq:gamma}
\end{equation}
since the number of loops of a certain length is $dn=dN/l$.
Both the scaling hypothesis and the assumption of Brownian statistics seem
to be good working hypotheses for $U(1)$ strings at formation.

\section{Low String Densities: the Hagedorn Transition}

Drawing on lessons learned from polymer statistics, the fact
that our algorithm generates Brownian strings is a result of the
dense packing of the strings. Naturally the question of what happens
for low string densities arises: we would like to find the
``natural'' dimension of a stringy random walk.  There is also
an important phase transition to investigate, that between a
percolating phase containing a certain fraction of infinite string,
and a low string density phase consisting of finite loops only.
In string theory
this is known as the Hagedorn transition \cite{Hag,BowCha,MitTur}.  It is
also of importance in spontaneously broken field theories with
vortex solutions \cite{Cop+}, for the stringy degrees of freedom
dominate the fluctuations near the phase transition, and generate
a divergence in the susceptibility slightly below the mean-field
critical temperature.  The question also arises of the order of the
transition. The calculations start from the assumption
that the string ensemble is Brownian, and then include the effect
of interactions between the strings in an approximate way.  They
find evidence that string-string interactions turn the
second-order transition obtained in the mean-field approach
into a first-order one.

Vachaspati \cite{V91} proposed an algorithm for reducing the string
density, which selects one of the three allowed field values, say 0,
as being more probable.  This reduces the probability of a string
penetrating the face of a lattice.  Thus we can generate an ensemble
with the {\it average\ } string density fixed at will.
(Physically one can think of this as applying an external field, which
spoils the $U(1)$ symmetry of the state.)
When the probability of string formation is made sufficiently
small, the strings stop
percolating. One also finds that the strings are more
crumpled, that is, they have dimension higher than 2.
Thus strings are not at all like polymers -- instead of being
self-avoiding they are `self-seeking' at low densities.
It is easy to convince oneself that
this is true by virtue of our lattice construction, and that this property
must increase with increasing $\eta$.

We introduce a bias parameter $\eta$ and set the
probabilities $p(k)$ of the vacuum phase $2k\pi/3$
as follows:
$p(0)=\eta/3$, $p(1)=p(2)=\frac{1}{2}(1-\eta/3)$. The probability
$P_{\rm f}$ for
a string to pass through a face is therefore $6p(0)p(1)p(2) =
\eta(1-\eta/3)^2/2$.  The probability $P_{\rm c}$ for a string to pass
through a cell is $2P_{\rm f}$, as the following argument makes clear.
If a string pass through a face, it must emerge through one of the
other faces of a tetrahedral cell with probability $1/3$.  Thus the
probability that a string passes through a particular pair of faces
is $P_{\rm f}/3$.  Since there are 6 pairs of faces, the total
probability is $2P_{\rm f}$.  Therefore the string line density is
\ben
\rho = \eta(1-\eta/3)^2
\label{eq:rhodef}
\een
in lattice units.  We can also express the expectation value of the
order parameter as a function of $\eta$:
\ben
\langle\phi\rangle = (\eta - 1)/2.
\een
Thus we see that, averaged over many lattice sites, the symmetry is
unbroken at $\eta=1$, moving towards $\langle\phi\rangle=1$ as $\eta\to 3$.

For low $\eta$,
the string network maintains a fraction of infinite string,
and the equations (\ref{eq:dimension})
and (\ref{eq:lengthdistr}) hold with $\eta$ dependent scaling exponents.
The equation which encodes the scale-free hypothesis for the loop
distribution, Eq.\ (\ref{eq:dn}), is not very well satisfied: the
scaling relation
$b=1+3/D$ is satisfied with errors of about 5\%, while the estimated
statistical error is just over 1\%.  It is, of course, possible that
we have underestimated our errors.
We record our measurements in the Table 1.
\vspace{5mm}
{\samepage
\begin{center}
\begin{tabular}{||l|l|l|l|l|l||}
\hline
$\eta$ & $N(l>25000)$ in \% & $D$ & $\rho$ & $b$ & $1+3/D$ \\
\hline
1.0  & 63.1$\pm$ 1.0  & 2.002$\pm$ 0.004 & 0.4444 & 2.49$\pm$0.02 & 2.50\\
1.1  & 59.5 $\pm$ 1.8 & 2.03$\pm$ 0.03   & 0.4412 & 2.47$\pm$0.03 & 2.48\\
1.2  & 44  $\pm$ 1    & 2.19$\pm$ 0.03   & 0.4320 & 2.32$\pm$0.03 & 2.37\\
1.21 & 41  $\pm$ 1    & 2.23$\pm$0.02    & 0.4308 & 2.22$\pm$0.02 & 2.35\\
1.22 & 38  $\pm$ 1    & 2.28$\pm$0.03    & 0.4295 & 2.20$\pm$0.02 & 2.32\\
1.23 & 33  $\pm$ 1    & 2.31$\pm$0.03    & 0.4282 & 2.18$\pm$0.02 & 2.30\\
1.24 & 28  $\pm$ 1    & 2.38$\pm$0.03    & 0.4268 & 2.14$\pm$0.02 & 2.26\\
1.25 & 25  $\pm$ 1    & 2.42$\pm$0.03    & 0.4253 & 2.11$\pm$0.02 & 2.24\\
1.26 & 19  $\pm$ 1    & 2.53$\pm$0.03    & 0.4238 & 2.09$\pm$0.02 & 2.19\\
1.27 & 14  $\pm$ 1    & 2.66$\pm$0.03    & 0.4223 & 2.05$\pm$0.02 & 2.13\\
1.28 & 10  $\pm$ 1    & 2.71$\pm$0.03    & 0.4208 & 2.04$\pm$0.02 & 2.11\\
\hline
\end{tabular}
\end{center}
{\footnotesize
{\bf Table 1 :} Scaling exponents and the string density (in units
of string elements per tetrahedron) for different values of $\eta<\eta_c$.}
}
\vspace{5mm}

These values up to $\eta\approx 1.25$ seem to all represent one
``high temperature''
phase of the network, and the statistics do not change qualitatively until
we reach the percolation threshold, where there is no longer any infinite
string.

In that domain Eq.\ (\ref{eq:lengthdistr}) has to be replaced by
Eq.\ (\ref{eq:lengthdistr2}), and scale invariance is no longer satisfied.
Instead, the parameter $c$ is the inverse of a diverging length scale, as it
appears in any second order phase transition. Besides the exponents in
Eq.\ (\ref{eq:lengthdistr2}) one can measure the susceptibility of the string
network as defined in Eq.\ (\ref{eq:S}) and the average length of string loops.
We list the measurements in the Table 2. We did observe evidence for
non-analyticity in the average
string length at the critical $\eta$, which may not be obvious from the
table below. Better statistics are needed to determine the (small and
positive) critical
exponent for $\langle\l\rangle$.
\vspace{5mm}
{\samepage
\begin{center}
\begin{tabular}{||l|l|l|l|l|l||}
\hline
$\eta$ & $\vev{l}$ & $\vev{l^2}$ & $S$ & $b$ & $c$
\\
\hline
2.2   & 10.4404 & 154.04  & 45.0386   & 1.635 & 0.0233    \\
2.1   & 11.649  & 207.831 & 72.1329   & 1.772 & 0.0157    \\
2.0   & 12.7826 & 273.722 & 110.329   & 1.687 & 0.0122    \\
1.9   & 14.3919 & 385.025 & 177.898   & 1.694 & 0.0089    \\
1.8   & 16.5119 & 584.274 & 311.633   & 1.776 & 0.00541   \\
1.7   & 18.8471 & 940.33  & 585.117   & 1.802 & 0.00286   \\
1.6   & 21.4212 & 1617.23 & 1158.36   & 1.838 & 0.00144   \\
1.55  & 23.3261 & 2358.77 & 1814.67   & 1.958 & 0.00104   \\
1.5   & 25.4684 & 3567.08 & 2918.44   & 1.989 & 0.00064   \\
1.45  & 27.932  & 5970.12 & 5189.9    & 1.977 & 0.00038   \\
1.4   & 30.5805 & 11442   & 10506.8   & 1.969 & 0.000193  \\
1.38  & 30.7761 & 14757.6 & 13810.4   & 1.998 & 0.000129  \\
1.37  & 32.1548 & 18380.6 & 17346.6   & 1.986 & 0.000111  \\
1.36  & 33.84   & 23280.3 & 22135.1   & 1.979 & 0.000091  \\
1.35  & 34.1428 & 28543   & 27377.2   & 2.004 & 0.000067  \\
1.34  & 34.0535 & 36636   & 35476.4   & 2.008 & 0.000048  \\
1.33  & 34.8022 & 47241.1 & 46029.9   & 2.025 & 0.000035  \\
1.32  & 34.9562 & 60152.3 & 58930.3   & 2.042 & 0.000024  \\
1.31  & 35.1275 & 79784.5 & 78550.5   & 2.045 & 0.000016  \\
1.30  & 35.9985 & 120865  & 119569    & 2.070 & 0.000009  \\
1.29  & 37.1514 & 200032  & 1986528   & 2.080 & 0.000005  \\
1.28  & 36.8624 & 348316  & 346957    & 2.085 & 0.000002  \\
\hline
\end{tabular}
\end{center}
{\footnotesize {\bf Table 2 :} The relevant measurements for $\eta>\eta_c$.
The divergence of the length scale
$1/c$ as well as the divergence of the susceptibility
indicate a second order phase transition at $\eta_c$.}
}
\vspace{5mm}

The assumption that the divergence of the susceptibility has a critical
exponent $\gamma$, so that
\begin{equation}
S=S_0(\eta-\eta_c)^{-\gamma},
\label{eq:sigma}
\end{equation}
and that similar exponents exist for the other variables in Table 2
gives a good estimate of the critical $\eta$. We measure
\[
\eta_c=1.243, \qquad
\gamma=2.36.
\]
There is also an exponent $\si$ for the loop size cut-off parameter $c$,
defined by
\ben
c \propto |\eta-\eta_c|^{\frac{1}{\si}}.
\een
The divergence of $1/c$ can be seen in Figure 7. A fit to this form
for $c$ gives
\ben
\eta_c=1.256 \qquad \si = 0.397.
\een
Thus we estimate $\eta_c = 1.25 \pm 0.01$.  The errors in the
exponents are more difficult to gauge, for they depend on the uncertainty
in $\eta$,  and we have not performed
a proper error analysis.  However, we believe that they can be trusted to
about 5\%.

\section{Comparison to Thermal Quantities}

If we define a probability density that a given string loop is in the
length interval $[l,l+dl]$ by $P(l)\propto{dn}/{dl}$ and the normalization
condition $\sum_l P(l)=1$, we can define an information entropy, contained
in the probability distribution function by
\[
S=-\sum_l P(l) \ln P(l)\mbox\ .
\]
The percolation transition that we observe is not a thermal phase transition,
but we can attempt to relate certain observables in our Monte Carlo ensemble
to variables in a canonical thermal ensemble.
We can for instance try to relate the average length of string loops to an
average energy of a thermal ensemble (our susceptibility in Eq.\
(\ref{eq:S}) relates then to a specific heat, which is indeed the case for
perfect $U(1)$ strings). It would then make sense to define a temperature
$\theta$ as
\[
\frac{1}{\theta}=\frac{\partial S}{\partial \langle l \rangle}\mbox\ ,
\]
and see how the system behaves under changes of this parameter.
Unfortunately we do not yet have sufficient statistics, since
quantities like $\langle l \rangle$ in Table 2 are not even monotonically
decreasing with increasing $\eta$. The necessity of improved
statistics is also apparent in Table 3, which lists the information entropy
of our Monte Carlo ensembles at different values of $\eta$. The finite
difference approximation to $\left[\partial S/ \partial \langle l \rangle
\right]^{-1}$ is statistically quite unstable because both $S$ and
$\langle l \rangle$ have small differentials compared to their absolute
values.  Improved statistics will be
obtainable with the algorithm under development, which will use a bigger
hashtable, with less hash-function collisions.
\vspace{5mm}
{\samepage
\begin{center}
\begin{tabular}{||l|l|l||}
\hline
 $\eta$   &       $S$ & $\langle l \rangle$ \\
\hline
1.28 &  1.19223 & 36.8624 \\
1.29 &  1.18510 & 37.1514 \\
1.3  &  1.16673 & 35.9985 \\
1.31 &  1.15198 & 35.1275 \\
1.32 &  1.14534 & 34.9562 \\
1.33 &  1.14681 & 34.8022 \\
1.34 &  1.14117 & 34.0535 \\
1.35 &  1.14373 & 34.1428 \\
1.36 &  1.14637 & 33.8400 \\
1.37 &  1.14292 & 32.1548 \\
1.38 &  1.14067 & 30.7761 \\
\hline
\end{tabular}
\end{center}
{\footnotesize {\bf Table 3 :} The values of the entropy and the average loop
length in our Monte Carlo ensembles. They are rather well defined,
although the ratio of their differentials
-- corresponding to a thermodynamic temperature -- has still very bad
statistical errors.}
}
\vspace{5mm}

\section{Conclusion}

We can apply our numerical technique to other
vacuum manifolds, whose non-contractible contours have a discrete mapping onto
triangular faces, such as the vacuum manifold of nematic liquid crystals
(NLCs),
which is $RP^2$, or
$SO(3)/O(2)$, the space points on a two-sphere with opposite points identified
(it is therefore also equivalent to $S^2/Z_2$). The statistics there could
be directly verified in the laboratory, and some variables, like the average
number of strings created per bubble after a first order phase transition into
the nematic phase, have in fact been measured already \cite{Bowick}.
The symmetry group of NLCs is very easily understood in terms of the
orientations
of the molecules. The symmetry manifold consists of the orientations of a
directionless rod. Many NLCs are polarisable, so applying a strong
electric field could make the
molecules prefer orientations parallel to the field axis. Thus one
could measure things like the percolation threshold and the
exponents such as $D$ and $b$.
Work in this direction is in progress.

We think that the numeric measurement of configurational statistics has never
been done with such accuracy, since the numerical methods used here allow
a much higher number of very large strings to be investigated,
without being influenced by boundary conditions. This gets rid of the
adverse effects of the topology of the lattice, and also speeds up the
computations considerably. We are able to confirm the scaling hypothesis
for the case of an exact $U(1)$ symmetry, and we find no compelling
evidence that the fractal dimension of the cosmic string is different
from 2.

As the string density decreases, we
measure an increase in the fractal dimension, which implies that the strings
get more crumpled and are statistically biased to turn back onto
themselves. This was indeed implicit in Vachaspati's results
\cite{V91}. Interestingly this effect has also been shown to occur
in the equilibriated phase of flat-spacetime string dynamics. In
ref.~\cite{SV} it is shown that for a sufficiently high initial string
density the string dynamics keep a certain fraction of string mass
alive in infinite strings, for reasons that lie beyond simple
flux-conservation arguments. Furthermore, ref.~\cite{SakV} confirms
that the Equations (\ref{eq:lengthdistr}) and
(\ref{eq:lengthdistr2}) hold for the high
and low density phase of string dynamics respectively, but with a
constant exponent of $b=\frac{5}{2}$ in both phases,
as one would expect for a thermalized ensemble \cite{MitTur}.
For a thermalized ensemble of string one
also expects the density in string loops to be constant
for all string densities above the critical density, as numerically
confirmed in ref.~\cite{SakV}. That this is not
the case for the cosmological initial conditions has been postulated
in \cite{V91} and had received further backing from our results. For
our static Monte Carlo ensemble the number of string loops
{\em increases} as we approach the critical string density from above
(see Table 1).

In the low density
phase, we were able to construct thermodynamic measures of the loop
statistics from their average length and the length fluctuations, related to
as energy and specific heat respectively.  The critical exponent
of the divergence of the susceptibility with temperature was -2.43.
The asymmetry parameter $\eta$ at the phase transition was $1.24\pm0.01$,
which corresponds to a vacuum expectation value
$\langle\phi\rangle=0.12$, and a string
density of $0.426$ strings per tetrahedron,
i.e.~$\langle length/volume \rangle=2.56/a^2$, where $a$ is the
edge length of the cubes of the body-centered cubic lattice, spanned
by the vertices of the tetrahedra, so that the correlation length
lies between $\sqrt{3}/2\,a$ and $a$. This value is considerably
higer than the one computed for differently constructed strings on
a cubic lattice in ref.~\cite{V91} ($\rho_c\approx 0.88/a^2$ there),
which might again partly be caused by the fact that we allow smaller
loops to form than there are on a cubic lattice. Nevertheless, one
does not expect the density to be universal for different ways of
constructing strings (i.e.\ for discretizations of different symmetry
groups), but further work on $RP^2$ is in progress and will shed more
light on this question. Also the question whether the critical
exponents exhibit universality with respect to the different ways
of discretizing the vacuum manifold and space, or even some universality
with respect to different vacuum manifolds, remains open.

We conclude with some discussion of the connections between our work and
other problems in statistical physics.  It is probably clear that what
we are considering is a kind of percolation problem \cite{Stauffer}. In
fact, our field of three phases is essentially the spin of a 3-state
Potts model, and the appearance of infinite string at the Hagedorn
transition can only happen when the disfavoured spins ($k=1,2$)
percolate. However, in conventional percolation theory one is generally
interested in the domain size distribution $n(s)$ and its moments: here,
we are picking out the junction between 3 domains for study.
Nevertheless, there are close parallels.  For example, the mean cluster
size is equivalent to $\vev{l^2}$, and the Fisher exponent $\tau$ for
the cluster size distribution, defined by $n(s) \propto
s^{-\tau}$ at criticality, is equivalent to our exponent $b$.  The
strength of the infinite cluster is related to the length in
infinite string, $\rho_\infty$. Indeed, if we define a new exponent
$\beta$ from the non-analytic behaviour of $\rho_\infty$ near the
transition, such that
\ben
\rho_\infty \propto |\eta-\eta_c|^\beta,
\een
these analogies suggest the following scaling relation:
\ben
\be = (b - 2)/\sigma.
\label{eq:beta}
\een
Another scaling relation should also hold:
\ben
\ga = (3 - b)/\sigma.
\een
Unfortunately, our statistics were not good enough to be able to
extract $\beta$ very satisfactorily.  From Table 1 it can
be seen that $\rho_\infty$ is still non-zero for densities
below the transition.  This is an effect of the length cut-off -- if
we could go to abitrarily long string, $\rho_\infty$ would presumably
vanish at $\eta_c$.
The scaling relation (\ref{eq:beta}) is also rather sensitive to values
of $b$ near 2.  The form of $\rho_\infty/\rho$, plotted in Figure 8,
is nevertheless indicative of a second order transition.

The second of the above scaling relations is easier to check:  with
$b(\eta_c)=2.11$ and $\si\simeq 0.397$, we should have $\ga=2.24$.  Our
measurements give $\ga=2.36$.  We do not have sufficient confidence
in our statistics to be able to say whether this is a significant
deviation, but it is encouraging to find  agreement to 5\%.

Our simulations also model the initial conditions of condensed matter
systems with a nonconserved order parameter after a rapid quench.
Dynamical simulations have been performed on systems where the order
parameter is a complex scalar field $\phi$, both with $\vev{\phi}=0$
(``critical'', corresponding to $\eta=1$), and $\vev{\phi}\ne 0$
(``off-critical'', corresponding to $1<\eta<3$) \cite{MonGol92}.  It is
found that the introduction of a bias in the initial expectation value
of the order parameter results in the eventual departure from dynamical
scaling, with the density of the string network going as
\ben
\rho(t) \sim t^{-1} \exp(-gt^{3/2}),
\een
where $g$ depends approximately quadratically on the initial bias
$\vev{\phi}$. This is due to the network breaking up into isolated
loops, with an exponentially suppressed size distribution.  What is not
clear is whether this is due to the initial conditions possessing no
infinite string, or whether the infinite string somehow manages to chop
itself up into an infinite number of loops.  If it is the former, then
that means that the percolation transition happens at very small bias,
perhaps even $\langle\phi\rangle=0$,
for the departure from power-law scaling in
$\rho(t)$ was observed from rather small biases, down to
$\langle\phi\rangle=0.001$.
This brings us back to the question raised earlier this section: how
does the discretisation affect the percolation transition?  We plan to
investigate this point further.

\def\rf#1#2#3#4#5{#1, {\it #3} {\bf #4}, #5 (#2)}

\def\APJ{Ap. J.}
\def\CMP{Comm. Math. Phys.}
\def\GRG{Gen. Rel. Grav.}
\def\JETP{Sov. Phys. JETP}
\def\JETPL{Sov. Phys. JETP Lett.}
\def\MNRAS{Mon. Not. R. Ast. Soc.}
\def\MP{Mod. Phys.}
\def\MPL{Mod. Phys. Lett.}
\def\YF{Yad. Fiz.}
\def\ZETF{Zh. Eksp. Teor. Fiz.}
\def\ZETFPR{Zh. Eksp. Teor. Fiz. Pis'ma Red.}

\def\AC{Acta Crystrallogr.}
\def\AM{Acta Metall.}
\def\AP{Ann.\ Phys., Lpz.}
\def\APNY{Ann.\ Phys., NY}
\def\APP{Ann.\ Phys., Paris}
\def\CJP{Can.\ J.\ Phys.}
\def\JAP{J.\ Appl.\ Phys.}
\def\JCP{J.\ Chem.\ Phys.}
\def\JJAP{Japan J.\ Appl.\ Phys.}
\def\JP{J.\ Physique}
\def\JPhCh{J.\ Phys.\ Chem.}
\def\JMMM{J.\ Magn.\ Magn.\ Mater.}
\def\JMP{J.\ Math.\ Phys.}
\def\JOSA{J.\ Opt.\ Soc.\ Am.}
\def\JPSJ{J.\ Phys.\ Soc.\ Japan}
\def\JQSRT{J.\ Quant.\ Spectrosc.\ Radiat.\ Transfer}
\def\NC{Nuovo Cim.}
\def\NIM{Nucl. Instrum. Methods}
\def\NP{Nucl.\ Phys.\ }
\def\PL{Phys.\ Lett.\ }
\def\PR{Phys.\ Rev.\ }
\def\PRL{Phys.\ Rev.\ Lett.\ }
\def\PRS{Proc.\ R.\ Soc.\ }
\def\PS{Phys.\ Scr.}
\def\PSS{Phys. Status Solidi}
\def\PTRS{Phil. Trans. R. Soc.}
\def\RMP{Rev.\ Mod.\ Phys.\ }
\def\RSI{Rev.i\ Sci.\ Instrum.\ }
\def\SSC{Solid State Commun.}
\def\ZP{Z.\ Phys.\ }

\renewcommand{\thesection}{}
\newpage
{\LARGE\bf Figure Captions}
\vspace{1cm}

{\bf Figure 1:} The tetrahedral subdivisions of a body-centered cubic lattice,
which do fulfill the requirements of having almost regular tetrahedra (i.e.\
edges of almost equal length) and of creating a string network which is in
a statistical sense isotropic, if the field values of the $U(1)$ manifold are
assigned to the vertices of the lattice.\vspace{1cm}

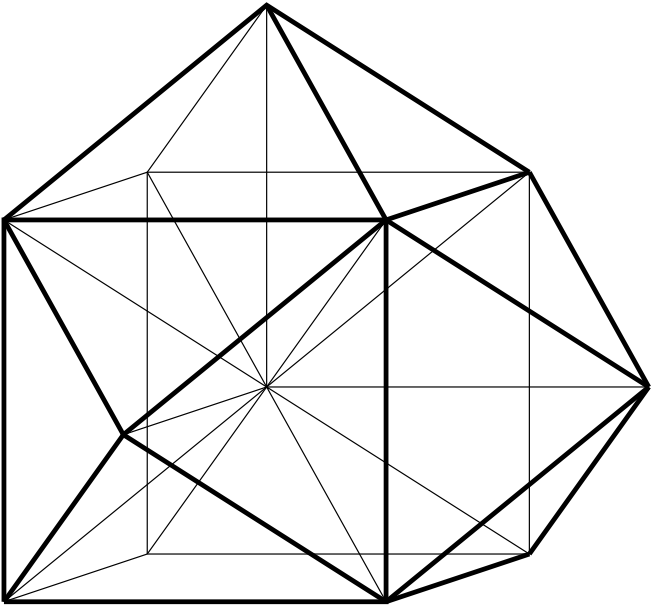
{\bf Figure 2:} The dual lattice of the lattice in Figure 1. The edges of this
lattice are the possible paths a string can take. Every vertex can be visited
only once, since it sits in the center of a tetrahedron of the lattice in
Figure one. This so-called tetrakaidekahedral lattice has the Brilloun zone
of the body-centered cubic lattice as its elementary cell. The field values
are assigned to the centers of these elementary cells. The near spherical shape
of the elemantary cell shows that this lattice is modelling the Kibble
mechanism rather well, since the border of this cell reprecents half the
distance to the event horizon.\vspace{1cm}

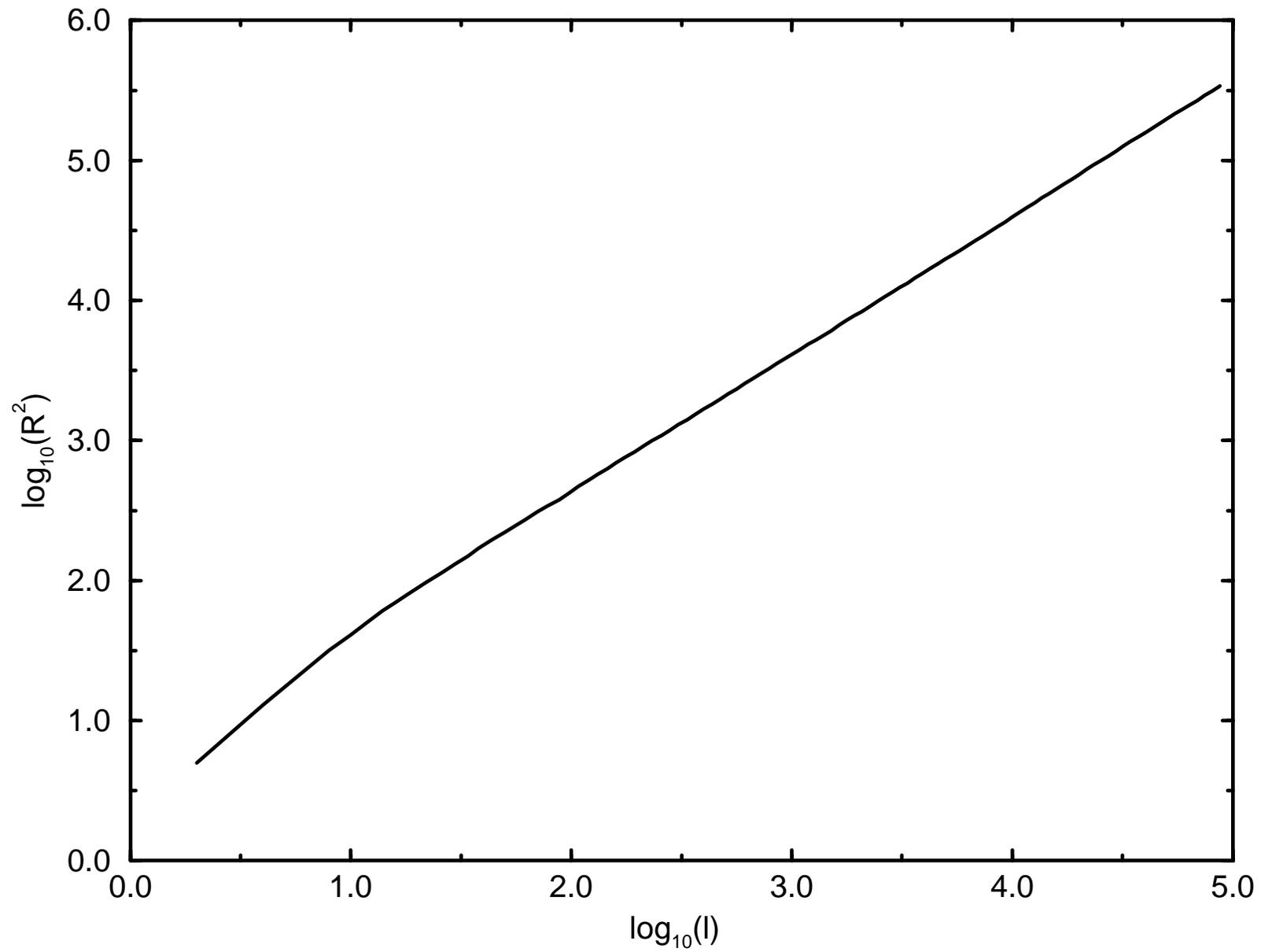
{\bf Figure 3:} The logarithmic plot of the square of the average distance
to the origin of a
point that is reached after having followed the string for a distance $l$. The
slope of the linear fit to this curve is $2/D$, with $D$ the
fractal dimension of the strings.\vspace{1cm}

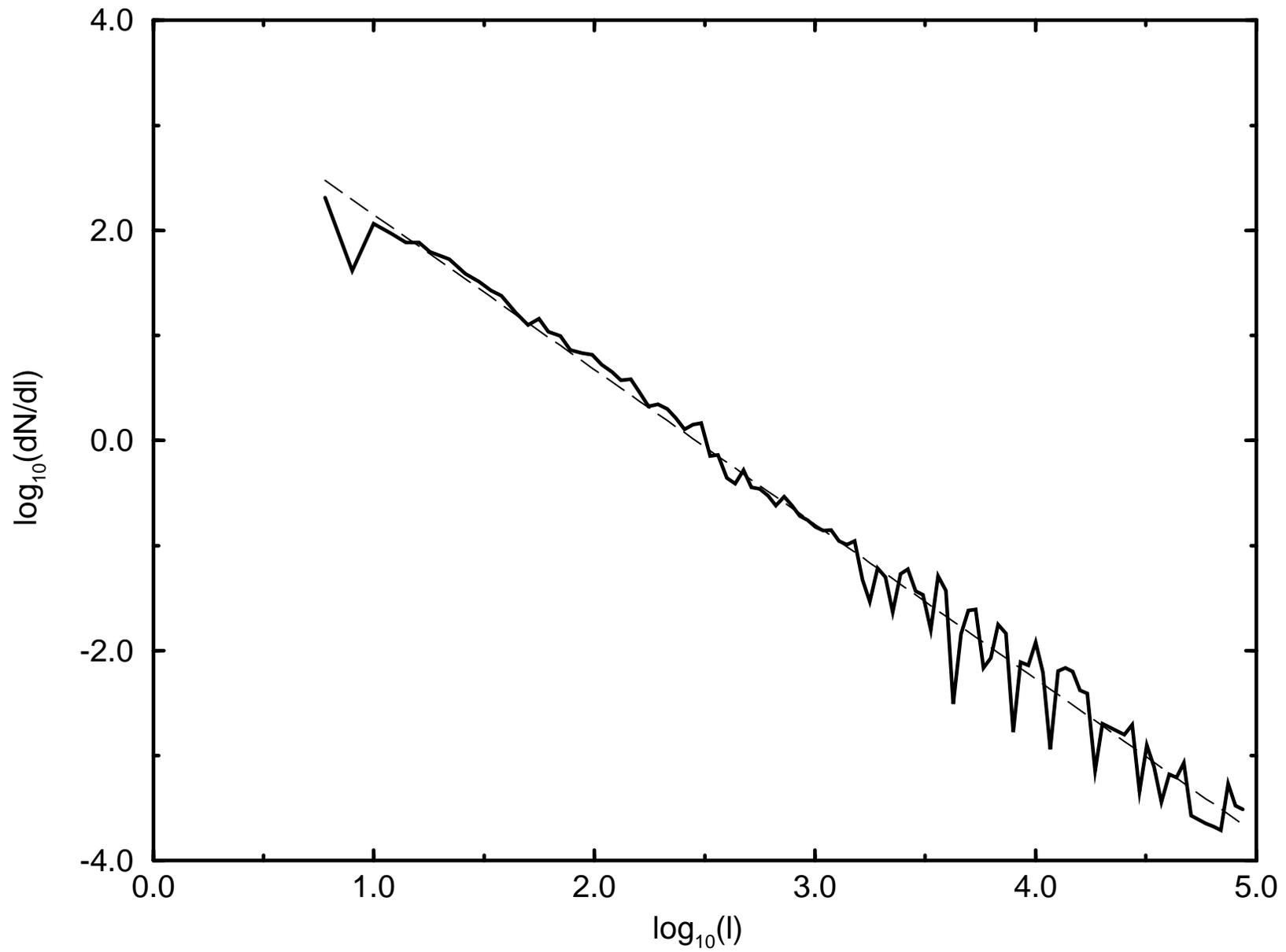
{\bf Figure 4:} Logarithmic plot of the number density of string mass
in loops within a certain length-range. The slope of the linear fit to this
gives one minus the exponent $b$ in Eq. (\ref{eq:lengthdistr}). We measure
$b=2.47\pm 0.02$, corresponding - if scaling is assumed - to
$D=2.04 \pm 0.03$, which can be taken as a consistency check for the scaling
assumption Eq.\ (\ref{eq:dimension}) rather than another convenient
way of measuring the exponent $D$.\vspace{1cm}

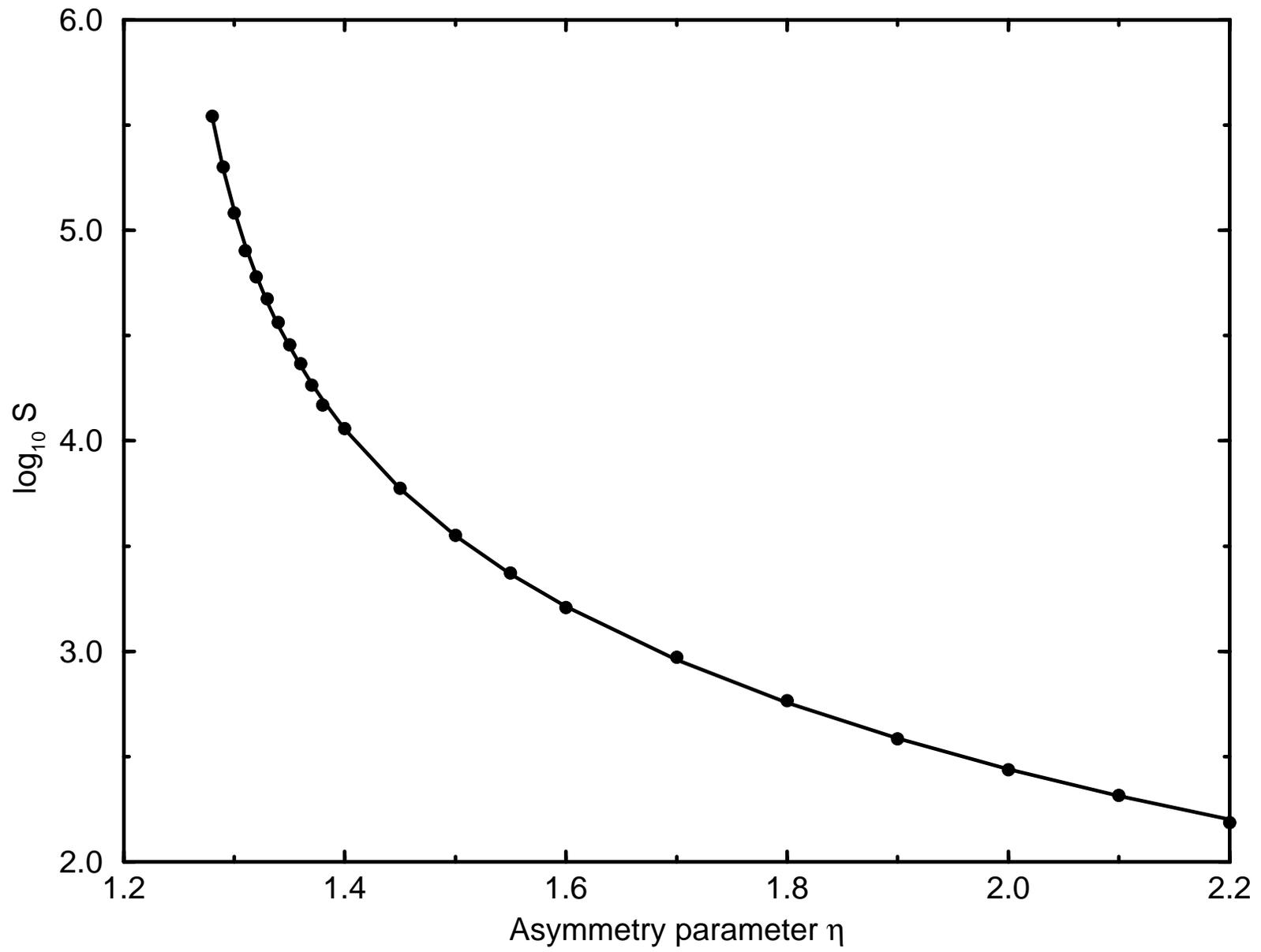
{\bf Figure 5:} The logarithm of the loop susceptibility $S$,
plotted against the asymmetry parameter $\eta$ (solid circles).
The line is a fit to the data, with $S=S_0(\eta-\eta_c)^
{-\gamma}$, where  $\gamma=2.36$ and $\eta_c = 1.243$.\vspace{1cm}

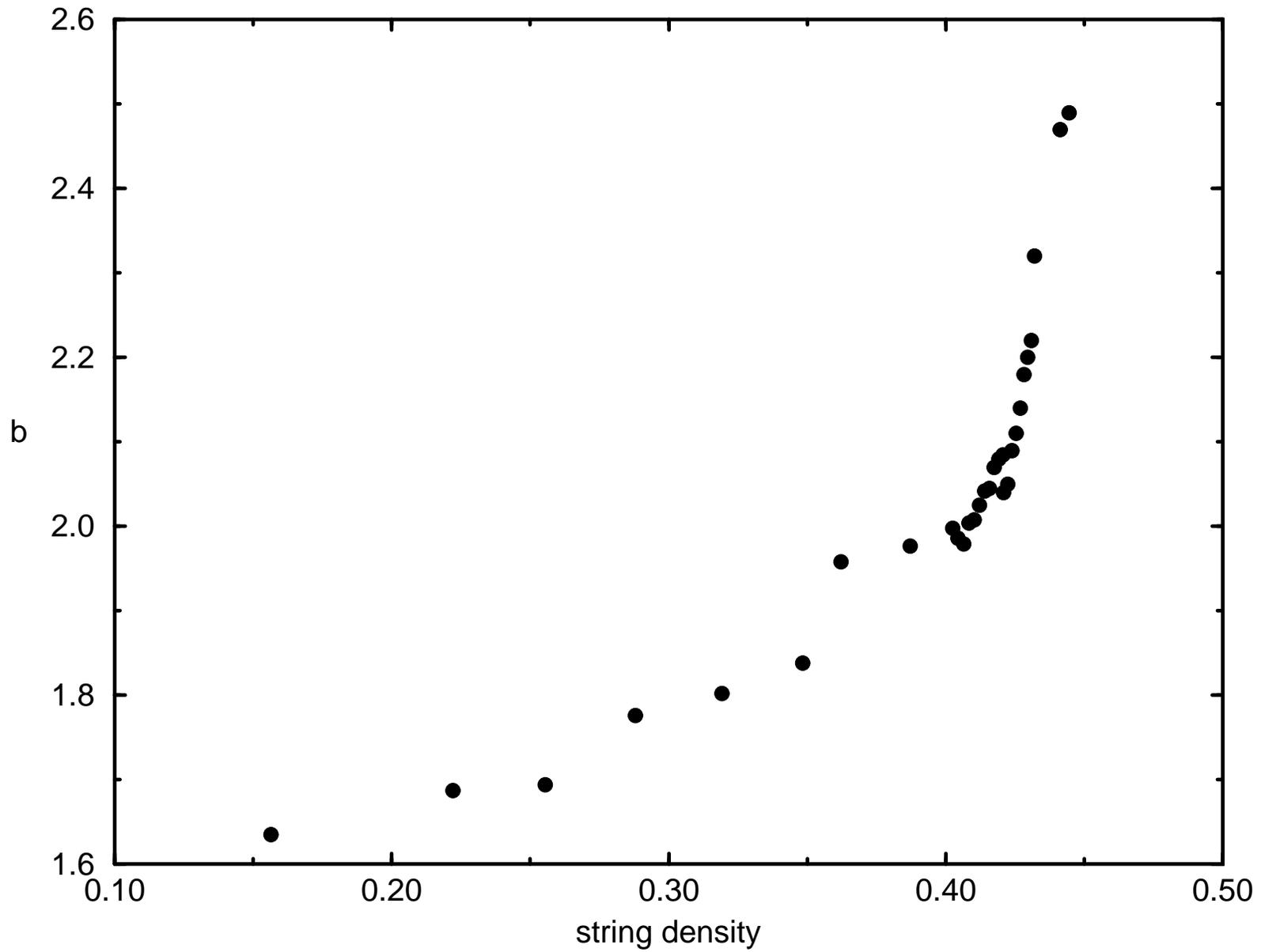
{\bf Figure 6:} A plot of the parameter $b$ in Eq.\ (\ref{eq:lengthdistr2})
vs.\ the density of strings. Vachaspati \cite{V91} stated that below the
percolation threshold of the string density the measurements of $b$ are
consistent with a value of 2. We don't find evidence for such behaviour,
but find that $b$ keeps decreasing monotonically with the string
density.

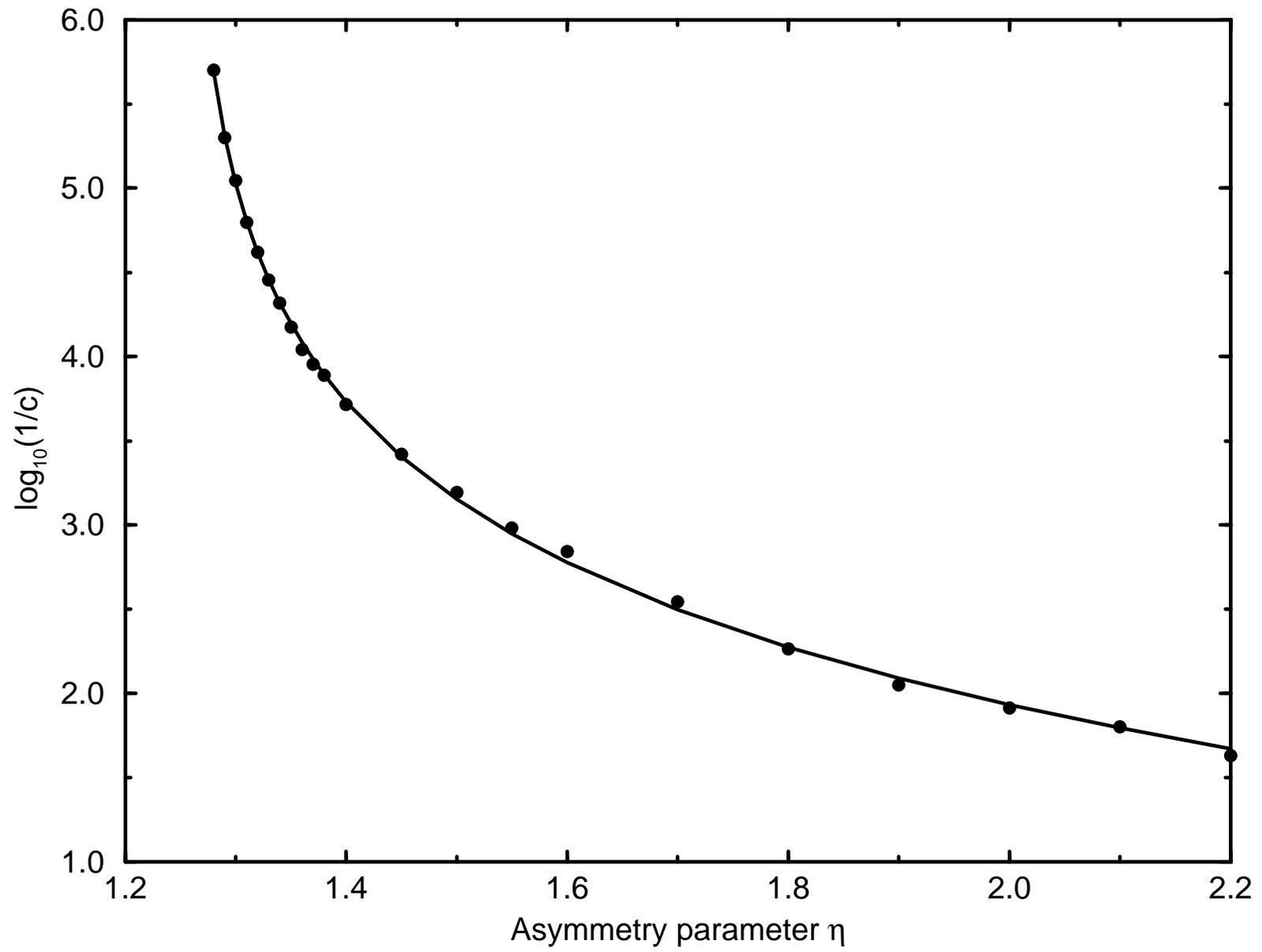
{\bf Figure 7:} The length scale on which the loop distribution is
exponentially suppressed for long loops diverges as $\rho\rightarrow
\rho_c -$. It is compared to the a fit of the form
$ c= c_0|\eta-\eta_c|^{1/\si}$, with $\eta_c=1.256$ and $\si=0.397$.
For $\eta>\eta_{c}$ there is no natural length scale
to be measured, and the scaling distribution Eq.\ (\ref{eq:lengthdistr})
is recovered.\vspace{1cm}

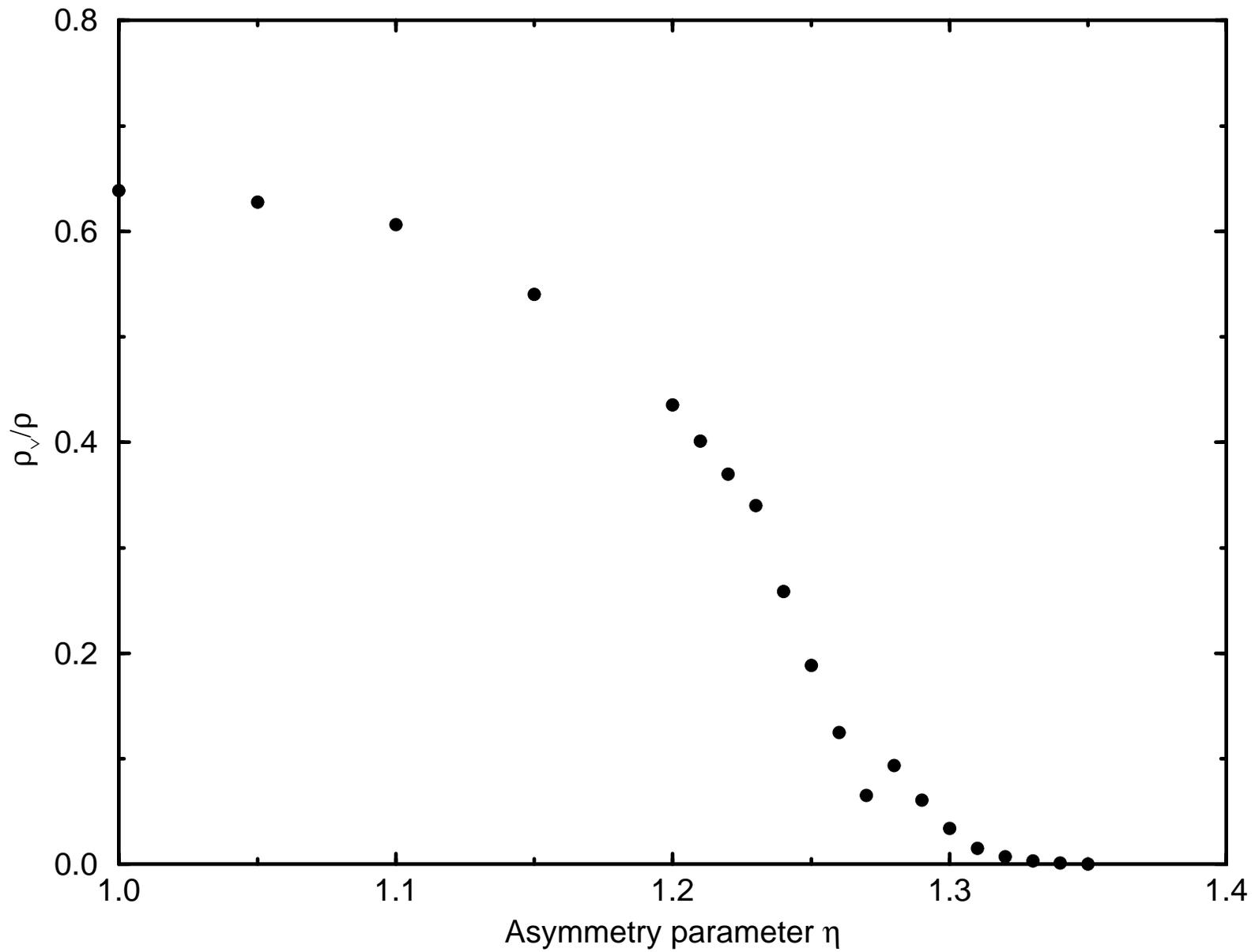
{\bf Figure 8:} The fraction of string which does not return to the
origin (``infinite'' string) $\rho_\infty/\rho$, where $\rho$ is
given by (\ref{eq:rhodef}), plotted against the asymmetry parameter.
This fraction is in effect an order parameter for the transition.

\end{document}